\documentclass[]{trbunofficial}
\usepackage{graphicx}
\usepackage[hidelinks]{hyperref}
\usepackage{url}
\usepackage{amsmath}
\usepackage{array}
\usepackage[table,xcdraw]{xcolor}
\usepackage{multirow}
\usepackage{colortbl}
\usepackage{makecell}
\usepackage{newtxtext,newtxmath}
\usepackage{geometry}
\usepackage{float}
\usepackage{todonotes}
\usepackage{subcaption}
\usepackage{rotating}
\usepackage{booktabs}
\usepackage{placeins}

\begin{document}

\AuthorHeaders{Ahmed, Hawkins, Armantalab}

\title{A Pseudo Panel Difference-in-Differences (DiD) Analysis of Online Shopping Behavior in the Puget Sound Regional Council (PSRC) Region}

\author{
  \textbf{Usman Ahmed}\\
  University of Tennessee - Oakridge Innovation Institute\\
  University of Tennessee - Knoxville\\
  Email: uahmed@utk.edu\\
  \hfill\break%
  \textbf{Jason Hawkins*}\\
  Department of Civil Engineering\\
  University of Calgary\\
  Email: jfhawkin@ucalgary.ca\\
  \hfill\break
  * Corresponding Author
  \hfill\break%
  \textbf{Omid Armantalab}\\
  Department of Civil and Environmental Engineering\\
  University of Nebraska Lincoln\\
  Email: oarmantalab2@huskers.unl.edu\\
  \hfill\break
}

\maketitle

\section{Abstract}
Online shopping is a growing trend, particularly following the COVID-19 lock-downs enacted by many cities. Understanding these trends requires robust panel dat methods. However, panel data (i.e., with repeated measurements of the same observational units) are often unavailable. In this study, we use a propensity score weighting (PSW) approach to adjust repeated cross-sectional travel diary surveys collected in the Puget Sound Regional Council (PSRC) region. The most common home delivery is packages (2.09 days per week in 2023), followed by food (0.36 days per week in 2023). We find that single-family detached and townhouse residents tend to receive more home deliveries than those living in apartments. We also find a difference in several patterns pre- and post-COVID pandemic. Vehicle deficient households did not exhibit the same increase in home delivery frequency as other households pre-COVID, but the pattern reversed in the post-COVID period - i.e., vehicle deficient households saw a relative increase in delivery frequency. Overall, this study demonstrates the pseudo-panel approach to causal inference by leveraging differences in PSW-weighted in delivery frequency across treatment groups.

\hfill\break%
\noindent\textbf{\textit{Keywords}: online shopping; pseudo-panel DiD; travel behavior and life events} 
\newpage

\section{Introduction}

The demand for online shopping has increased substantially over the past decade, driven by the rapid growth of e-commerce platforms and further accelerated by the COVID-19 pandemic. In the US, the share of e-commerce retail sales as the percent of total retail sales has increased to about 16\% in the first quarter of 2025 from around 7.8\% in the first quarter of 2016 \cite{retail2025}. Consumer preferences are also evolving, with increasing demand for flexible delivery options that balance speed, cost, and sustainability \cite{consumer2025ecommerce}. The demand for fast and same-day deliveries is also increasing which results in frequent deliveries with smaller shipment size. These shifting preferences are reshaping last-mile logistics strategies, such as proximity logistics, micro-distribution centers, and micro-delivery freight vehicles such as electric-cargo tricycles. Understanding these evolving consumer priorities is essential for designing responsive and resilient e-commerce and transportation systems. Much of the existing literature on online shopping demand focuses on its relationship with in-store shopping, revealing mixed findings. While some studies suggest a substitution effect \cite{shi2019does,xi2020impacts}, whereby online shopping reduces the frequency or duration of in-store visits, others point to a complementary relationship, whereby digital browsing leads to an increase in in-store shopping \cite{aldo2023exploring,cao2012interactions,zhen2016associations}. These differing results highlight the complexity of consumer behavior and other contextual factors that impact online shopping demand. Studies have examined various factors influencing online shopping demand including socio-demographic characteristics such as education level, internet usage, income, vehicle ownership, household size, and land use related variables such as population density. However, it remains unclear whether changes in these factors over time would lead to corresponding changes in online shopping behavior. Our study aims to fill this research gap.

With online shopping becoming more common, there has been a corresponding increase in publications focused on its implications for transportation planning and operations. Most studies rely on cross-sectional survey instruments to determine trip rates by purchase category  \cite{Pernot_2021,Hood_Urquhart_Newing_Heppenstall_2020,Lee_Sener_Handy_2015,Cao_Xu_Douma_2012,Rotem-Mindali_Weltevreden_2013}. \citep{Tejada_Conway_2025} used the 2019 Citywide Mobility Survey (CMS) collected by New York City Department of Transportation to model online shopping as a binary process of whether a person received a delivery on each of seven days. The CMS categorizes deliveries as one of groceries, prepared food, or other packages. Their focus was the effect of demographics and land use on home delivery frequency. They found that grocery delivery was more common in high-density areas, while prepared food delivery was most common in low-density areas. \citep{Figliozzi_Unnikrishnan_2021} conducted a survey in the Portland metropolitan area in May and June 2020 to capture the influence of the COVID-19 pandemic on home delivery patterns. They used choice models to examine order frequency and/or expenditure, latent variables, attitudinal/ordered responses, and socioeconomic factors, as well as public health concerns. Their focus was equity through a home-based accessibility measure.  \citep{Figliozzi_Unnikrishnan_2021} found that underserved populations were also less likely to benefit from home delivery service improvements during this period. \citep{Newing_Hood_Videira_Lewis_2022} examined a similar question of low online grocery delivery access within rural areas of the UK. \citep{Beckers_Cárdenas_Verhetsel_2018} used an e-commerce survey conducted by the Belgian retail federation Comeos to investigate general package delivery rates in Belgium as a function of socioeconomic characteristics. They applied logistic regression models to examine the geographic distribution of demand for package deliveries. In addition, the authors point out that the level of urbanization does not influence the decision to shop online.

Several studies were conducted during 2020-2021 of the effect of the COVID-19 pandemic on online shopping and home delivery patterns \citep{Zhang_Ji_Lv_Ma_Kuai_Li_2022}. Wang et al. \cite{Wang_Gao_Nurul_Habib_2024,Wang_Gao_Liu_Nurul_Habib_2023} used survey data collected in Toronto, Canada to model the relationship between online and in-person grocery shopping. They found that the cost of home delivery acted as a barrier to online shopping with home delivery. They also considered online ordering and pick up, finding that such shoppers are twice as sensitive to travel time when compared with in-store shoppers. \citep{Meister_Winkler_Schmid_Axhausen_2023} find a similar preference for in-person grocery shopping in Switzerland during the same period. A  study based on a US sample survey found that work-from-home frequency is positively correlated with online non-grocery shopping \cite{Mohammadi_Davatgari_Asgharpour_Shabanpour_Mohammadian_Derrible_Pendyala_Salon_2024}. These studies anticipated that online grocery shopping would increase post-COVID, but there has been minimal follow-up analysis to validate these predictions.

Life events such as household relocation and vehicle ownership influence individual behavior. For example, studies have found that life events affect travel behavior of individuals \cite{ahmed2023impact,clark2014life}. The impact of life events on changes in online shopping demand remains under-researched, despite its growing impacts on travel behavior and urban goods movements. Analyzing such effects typically requires panel survey data which is often not available and expensive to conduct. Most of the travel surveys in the United States are cross-sectional surveys and are periodically conducted. These surveys are often available publicly, such as the National Household Travel Survey \cite{nhts2022}. Such surveys are useful for modeling online shopping demand for a single year but cannot capture changes in online shopping demand due to changes in life events. In this study, we introduce a novel approach that leverages repeated cross-sectional travel surveys to examine how changes in life events impact online shopping demand. Specifically, we introduce a pseudo-panel framework, which constructs cohorts based on shared characteristics across survey waves, allowing us to analyze temporal dynamics without relying on costly panel data. This approach provides valuable insights into the evolving relationship between life transitions and online shopping demand. In this study, we focus on three types of life events: 1) household tenure (rent vs. owning), 2) dwelling type, and 3) vehicle ownership. We also explore the impact of COVD-19 on changes in online shopping demand.

Beginning with the 2003 study by \citep{Anderson_Chatterjee_Lakshmanan_2003}, there are two standard hypotheses about the effect of residential location on online shopping frequency. The first is that urban residents are early adopters of technology and more likely to adopt online shopping (innovation-diffusion hypothesis). The second hypothesis is that the lower access to in-person shopping opportunities among suburban and periurban residents leads to more shopping, as these residents can access a wider diversity of shopping options online (efficiency hypothesis). Results as to which of the two hypotheses govern are mixed, with variation across studies conducted in different regions \cite{Farag_Schwanen_Dijst_Faber_2007,Chevallier_Coninck_Motte_Baumvol_2016,Weltevreden_2007}. Individuals often consult online before making an in-person purchase, a consistent finding among studies over the last 15 years of market evolution \cite{Weltevreden_2007,Etminani_Ghasrodashti_Hamidi_2020}. \citep{Etminani_Ghasrodashti_Hamidi_2020} find that time pressure is a greater determinant of online shopping than technology enthusiasm. They also emphasize the importance of considering variation across goods types.

We found that most studies take a cross-sectional approach to e-shopping behavior, missing the dynamics of this rapidly evolving market. We contribute to the literature by applying a robust pseudo-panel weighting approach to consider e-shopping dynamics for a large metropolitan area in the US. We take advantage of survey data collected pre, during, and post COVID-19 to measure its effect on e-shopping behavior. We also consider how these patterns are affected by changes in housing choice. In the next section, we outline our econometric modeling approach. We then describe the dataset used in the empirical study. Model results are presented and conclusions drawn. The paper ends with a discussion of future research directions.

\section{Methods of Analysis}
Inference on behavioral evolution requires data collected at multiple time points (i.e., longitudinal data). Ideally, one would collect longitudinal panel data that maintains the same sample across time points \cite{Frethey_Bentham_2011}. However, panel data is more costly and time intensive to collect. It also does not capture population and average values at distinct points as well as cross-sectional data \cite{Frethey_Bentham_2011}. Panel attribution also means that the effective sample size of consistent observational units over the panel period can be low \cite{Murakami_Ulberg_1997,Yee_Niemeier_1996}. As such, it is typical for transportation planning agencies to structure their travel diaries as cross-sectional samples. The question that arises for the researcher is how to leverage such data for longitudinal analysis. We identified two methods of pseudo-panel construction in the literature. 

The first approach, attributed to \citep{Deaton_1985} aggregates data into cohorts based on demographic variables that are common across years. The mean variable value is used for a continuous outcome variable, while either the mean or modal value is used for a categorical outcome variable. The notion behind this approach is that aggregation into sufficient large cohorts will result in reasonable approximations of population statistics, assuming cross-sections represent random samples from each cohort in success years. The advantage is that one can directly compare cohort outcomes over time. However, statistical efficient is diminished by aggregation because it reduces the sample size. Also, aggregation may result in the loss of other variables not included in the cohort variable set. This first approach appears to be the more common of the two in the literature \cite{Guillerm_2017,Dargay_2002,Matas_Raymond_2008,Weis_Axhausen_2009,Huang_2007}.

The second approach uses matching to fuse datasets collected at different time points. The conventional fusion application involves two datasets containing different variables for the same observations, as well as a subset of common variables. The fusion is accomplished by first defining donor and recipient datasets, then matching the common variables across the two datasets. It is typical in the case of continuous variables to use a nearest neighbor heuristic. However, the fusion process becomes more complicated when one considers two time points as the fusion variables may change between collection periods. E.g., for an individual who has age $a_t$ at time $T$, one must find an individual with age $a_t+1$ at time point $T+1$ years. The process becomes even more complex when one adds time periods to the fused dataset. Unlike the first approach, fusion maintains the sample size and all variables in the final dataset. However, estimation efficiency will degrade if matching variables are imbalanced and it may become invalid to directly compare observations across time periods. \citep{Frethey_Bentham_2011} proposed but did not implement the approach and we were unable to find an application in subsequent work citing the paper. This setup is similar to the two-way PSM quasi-experimental approach of \citep{Zhong_et_al_2021_two_dimensional} in the transportation literature.

Within the econometric literature, this second approach based on data fusion can be considered as a form of PSM. The assumption made in PSM is that the outcome variables $Y_t$ and $Y_{t+1}$ at the two time points are conditionally independent given the matching variables. A related approach is propensity score weighting (PSW), which operates in the same way but does not require a direct mapping of observations between years. Like survey weights to adjust a sample to the population demographic distributions, PSW can be used to adjust for differences in the cross-sectional samples relative to a reference sample. For the PSW approach, covariate balancing variables are used to estimate the propensity of an observation being drawn from a given time point and treatment condition assuming a multinomial discrete outcome.  It is based on the propensity score difference-in-differences (DiD) modelling approach proposed by \citep{Stuart_Huskamp_Duckworth_Simmons_Song_Chernew_Barry_2014}. They show that outcome regressions weighted by this PSW can provide causal parameter estimates with suitable caution when defining the covariates for the PSW function. We used the WeightIt package in R \cite{Greiger_2025_WeightIt} to test several weighting algorithms. The tested algorithms were Bayesian Additive Regression Trees (BART), Gradient Boosted Machines (GBM), SuperLearner, and multinomial logistic regression. Other algorithms are available in WeightIt but are not applicable to our problem.

Ordered probit models were estimated for the outcome of total days with a home delivery. We tested three models considering change in tenure and building type in the prior five years as sources of variation. The models for tenure change are given below (dwelling type models have parallel structures, with \emph{own} being replaced by \emph{detached/semi-detached dwelling}

\begin{equation}\label{eq:model_1}
y_i = \beta_{0i} + \beta_{own} (Own=1) + \beta_{year} (Year = T_1) + \beta_{own,year} ((Own=1) \times (Year=T_1))
\end{equation}
\begin{equation}\label{eq:model_2}
y_i = \beta_{0i} + \beta_{tenure} (Tenure_0 \neq Tenure_1) + \beta_{R2O} (Tenure_0 = Rent \text{ \& } Tenure_1=Own)
\end{equation}
\begin{equation}\label{eq:model_3}
\begin{aligned}
y_i =\; & \beta_{0i} 
+ \beta_{own} (Own = 1) 
+ \beta_{year} (Year = T_1)  + \beta_{own,year} \big((Own = 1) \times (Year = T_1)\big) \\
& + \beta_{R2O} (Tenure_0 = Rent \text{ \& } Tenure_1 = Own) \\
& + \beta_{R2O,year} \big((Tenure_0 = Rent \text{ \& } Tenure_1 = Own) \times (Year = T_1)\big)
\end{aligned}
\end{equation}
\\
where $y_i$ is the ordered outcome for delivery frequency in days per week; $T_0$ and $T_1$ are the initial and final year, respectively; and $Tenure_0$ and $Tenure_1$ are the initial and final dwelling tenure following a move, respectively. Model \ref{eq:model_1} uses a combined propensity score and survey weight, while Models \ref{eq:model_2} and \ref{eq:model_3} use the survey weight only because they use samples filtered for households that moved in the prior five years.

An additional model was estimated for each pair of years as follows

\begin{equation}\label{eq:model_4}
\begin{split}
   y_i = \beta_{0i} + \beta_{veh} (VehicleDeficient=1) + \beta_{year} (Year = T_1) + \\ 
   \beta{veh,year} ((VehicleDeficient=1) \times (Year=T_1)) 
\end{split}
\end{equation}

\section{Data}

The dataset for this research was the Puget Sound Regional Council (PSRC) household travel survey. The survey is conducted every two years. We used data for 2017, 2019, and 2023 to contrast trends for pre- and post-COVID periods. The survey includes information on current and previous residential locations. Interestingly, PSRC ran a panel travel survey (the first in the United States) between 1989 and 1993 \cite{Yee_Niemeier_1996} but discontinued it. The PSRC travel survey collected in 2019 was previously used by \citep{Shah_Carrel_Le_2024} in a study of online shopping, telework, and their influence on household tour rates and composition. They found that mandatory tours were negatively associated with online shopping but no statistically significant relationship between online shopping and other tour types.

\citep{Dias_Lavieri_Sharda_Khoeini_Bhat_Pendyala_Pinjari_Ramadurai_Srinivasan_2020} used the 2017 PSRC survey to study the factors that influence online and in-person activity engagement. They estimated multivariate ordered probit models of multiple activity engagement probability and frequency. The authors found a complementary relationship between in-person and online non-grocery shopping, but a substitution relationship for in-person and online grocery shopping. Related to our focus on the residential context, they found that households in denser urban areas engaged in more frequent online shopping than their non-urban counterparts.

Table ~\ref{tab:desc_stats} provides a summary of reported average delivery frequency by category and year. There has been a steady increase in delivery across all categories. The most common delivery is packages (more than 80\% of all deliveries). Food deliveries were constant prior to the COVID-19 pandemic but rose in 2023. The only category to decrease was services, which rose in 2019 but fell to 2017 levels again in 2023.

\begin{table}[h]
\centering
\caption{Average Delivery Frequency by Category (days per week) }
\label{tab:desc_stats}
\begin{tabular}{@{}cccccc@{}}
\toprule
\textbf{Year} & \textbf{Food} & \textbf{Services} & \textbf{Grocery} & \textbf{Packages} & \textbf{All} \\ \midrule
2017          & 0.23          & 0.25              & 0.20             & 1.64              & 2.04         \\
2019          & 0.23          & 0.31              & 0.19             & 1.94              & 2.37         \\
2023          & 0.36          & 0.24              & 0.20             & 2.09              & 2.50         \\ \bottomrule
\end{tabular}
\end{table}

\section{Results}
\subsection{Propensity Score Weighting (PSW) Model Results}
The PSW algorithm used in final ordered probit model estimation was determined based on the following weight statistics: coefficient of variation, maximum absolute deviation (MAD), entropy, effective sample size, and extreme weight magnitudes for combinations of observation year and treatment variable level. Results are provided in Tables \ref{tab:weightit_results_tenure} and \ref{tab:weightit_results_bart} (excluding SuperLearner, which gave consistently extreme weights). Tables \ref{tab:weightit_results_tenure} shows the results for all algorithms for tenure as the treatment and the 2017/2019 comparison. For brevity, we show only the BART results for all treatments in Table \ref{tab:weightit_results_bart}, which consistently provided the best results. We used household size, income, lifecycle stage, home county, and vehicle count as the covariate balance variables for all models based on a comparison of results for the BART algorithm using tenure as the treatment and the 2017/2019 comparison.

\begin{table}[H]
\centering
\caption{Full PSW Results for Tenure Treatment Variable (2017-2019)}
\label{tab:weightit_results_tenure}
\begin{tabular}{@{}lcccc@{}}
\toprule
                     & \textbf{2017-Rent} & \textbf{2017-Own} & \textbf{2019-Rent} & \textbf{2019-Own} \\ \midrule
 & \multicolumn{4}{c}{\textbf{COV}}                                                                                                                                         \\ \midrule
BART                 & 0.581                                  & 0.896                                 & 1.132                                  & 2.092                                 \\
GBM                  & 2.387                                  & 1.801                                 & 2.524                                  & 2.509                                 \\
GLM                  & 2.955                                  & 3.549                                 & 13.382                                 & 3.181                                 \\ \midrule
                     & \multicolumn{4}{c}{\textbf{MAD}}                                                                                                                                         \\ \midrule
BART                 & 0.397                                  & 0.547                                 & 0.466                                  & 0.654                                 \\
GBM                  & 1.337                                  & 1.172                                 & 1.357                                  & 1.225                                 \\
GLM                  & 1.412                                  & 1.341                                 & 1.638                                  & 1.508                                 \\ \midrule
                     & \multicolumn{4}{c}{\textbf{Entropy}}                                                                                                                                     \\ \midrule
BART                 & 0.129              & 0.261            & 0.273              & 0.528             \\
GBM                  & 1.376              & 1.029             & 1.465              & 1.268            \\
GLM                  & \multicolumn{1}{c}{1.667}              & 1.643             & 3.805              & 1.508             \\ \midrule
                     & \multicolumn{4}{c}{\textbf{ESS (unweighted/weighted)}}                                                                                                                   \\ \midrule
BART                 & 659/493                                & 534/296                               & 931/408                                & 984/183                               \\
GBM                  & 49/99                                  & 70/126                                & 87/126                                 & 148/135                               \\
GLM                  & 49/68                                  & 70/39                                 & 87/5                                   & 148/89        \\      \bottomrule                 
\end{tabular}
\end{table}

\begin{table}[t!]
\centering
\caption{Full PSW Results for All Treatment Variables Using BART Algorithm}
\label{tab:weightit_results_bart}
\begin{tabular}{@{}lcccc@{}}
\toprule
                    & \textbf{COV} & \textbf{MAD} & \textbf{Entropy} & \textbf{ESS (unweighted/weighted)} \\ \midrule
\textbf{2017-Rent}  & 0.671                            & 0.417                            & 0.158                                & 569/455                                                \\
\textbf{2017-Own}   & 0.887                            & 0.543                            & 0.256                                & 534/299                                                \\
\textbf{2019-Rent}  & 1.297                            & 0.469                            & 0.293                                & 931/347                                                \\
\textbf{2019-Own}   & 1.297                            & 0.469                            & 0.293                                & 984/199                                                \\ \midrule
\textbf{2019-Rent}  & 3.231                            & 0.641                            & 0.696                                & 931/81                                                 \\
\textbf{2019-Own}   & 1.109                            & 0.493                            & 0.269                                & 984/441                                                \\
\textbf{2023-Rent}  & 0.693                            & 0.452                            & 0.179                                & 574/388                                                \\
\textbf{2023-Own}   & 1.168                            & 0.562                            & 0.323                                & 700/296                                                \\ \midrule
\textbf{2017-Other} & 0.413                            & 0.298                            & 0.073                                & 472/403                                                \\
\textbf{2017-SF/TH} & 0.234                            & 0.162                            & 0.025                                & 738/700                                                \\
\textbf{2019-Other} & 0.293                            & 0.218                            & 0.038                                & 965/889                                                \\
\textbf{2019-SF/TH} & 0.206                            & 0.150                            & 0.020                                & 982/942                                                \\ \midrule
\textbf{2019-Other} & 0.367                            & 0.236                            & 0.053                                & 965/850                                                \\
\textbf{2019-SF/TH} & 0.270                            & 0.170                            & 0.030                                & 982/915                                                \\
\textbf{2023-Other} & 0.406                            & 0.293                            & 0.070                                & 749/643                                                \\
\textbf{2023-SF/TH} & 0.454                            & 0.286                            & 0.084                                & 568/471                                                \\ \midrule
\textbf{2017-VD}    & 0.452                            & 0.299                            & 0.084                                & 693/575                                                \\
\textbf{2017-NotVD} & 0.718                            & 0.446                            & 0.178                                & 1066/718                                               \\
\textbf{2019-VD}    & 0.697                            & 0.441                            & 0.179                                & 1066/718                                               \\
\textbf{2019-NotVD} & 0.504                            & 0.373                            & 0.106                                & 881/703                                                \\ \midrule
\textbf{2019-VD}    & 0.466                            & 0.324                            & 0.092                                & 1066/876                                               \\
\textbf{2019-NotVD} & 0.715                            & 0.470                            & 0.183                                & 881/583                                                \\
\textbf{2023-VD}    & 0.709                            & 0.437                            & 0.182                                & 852/567                                                \\
\textbf{2023-NotVD} & 0.826                            & 0.520                            & 0.244                                & 465/277  \\ \bottomrule
\end{tabular}
\caption*{\footnotesize SF = Single family detached; TH = Townhouse; VD = Vehicle deficient.}
\end{table}

\FloatBarrier

\subsection{Difference-In-Difference (DiD) Ordered Logit Model Results}

We estimate three model specifications using an ordered probit link function for total shopping frequency in days per week. The first model specification considers household tenure and its change between years. The second and third specifications use the subset of data for households that changed tenure from their last dwelling (i.e., rent to own or vice versa). Based on AIC statistics and for brevity, we present results for the first model specification only. Table \ref{tab:model_results_tenure} shows the results for the model using tenure as the treatment variable. Households that own their dwelling were less likely to receive deliveries during the 2017-2019 period, whereas they were more likely to receive them during the period 2019-2023. Examining the interaction term, we observe an increase in deliveries by owners between 2017 and 2019 but a slight decrease in 2023. Considering the presence of the COVID-19 pandemic in 2020-2021, it is intuitive that overall deliveries increased during this period (Year = $T_1$ parameter for the 2019-2023 model). 

\begin{table}[h]
\centering
\caption{Ordered probit model results for shopping frequency considering dwelling tenure}
\label{tab:model_results_tenure}
\begin{tabular}{@{}lcc@{}}
\toprule
\textbf{Variable}          & \textbf{2017-2019 Estimate (t-value)} & \textbf{2019-2023 Estimate (t-value)} \\ \midrule
Own property               & -0.31 (-55.3)                         & 0.20 (61.5)                           \\
(Year = $T_1$)               & -0.072 (-11.9)                        & 0.26 (82.5)                           \\
Own property $\times$ (Year = $T_1$) & 0.46 (62.6)                           & -0.080 (-19.8)                        \\
0|1                        & -0.64 (-132.7)                        & -0.69 (-252.0)                        \\
1|2                        & -0.13 (-27.3)                         & -0.021 (-7.70)                        \\
2|3                        & 0.32 (66.9)                           & 0.43 (157.8)                          \\
3|4                        & 0.54 (112.4)                          & 0.78 (282.2)                          \\
4|5                        & 0.98 (197.0)                          & 1.20 (425.8)                          \\
\midrule
Residual deviance          & 1,324,376                             & 4,268,275                             \\
AIC                        & 1,324,392                             & 4,268,291                             \\ \bottomrule
\end{tabular}
\end{table}

Table \ref{tab:model_results_dwelling} shows the results for the model using dwelling type as the treatment variable. It is consistently found that apartment residents receive fewer deliveries than single-family detached and townhouse (SFDTH) residents . Considering the interaction terms, it is found that apartment residents saw a decline in delivery rates in the period 2017-2019 relative to SFDTH residents but a relative increase in the period 2019-2023.
\begin{table}[h]
\centering
\caption{Ordered probit model results for shopping frequency considering dwelling type}
\label{tab:model_results_dwelling}
\begin{tabular}{@{}lcc@{}}
\toprule
\textbf{Variable}          & \textbf{2017-2019 Estimate (t-value)} & \textbf{2019-2023 Estimate (t-value)} \\ \midrule
Apartment/other               & -0.66 (-14.8)                         & -0.73 (-9.4)                           \\
(Year = $T_1$)                & 0.25 (4.6)                        & 0.36 (3.7)                           \\
Apartment/other $\times$ (Year = $T_1$) & -0.23 (4.1)                           & 0.11 (1.1)                        \\
0|1                        & -0.99 (-22.4)                        & -1.37 (-17.7)                        \\
1|2                        & -0.27 (-6.1)                         & -0.38 (-4.9)                        \\
2|3                        & 0.36 (8.2)                           & 0.44 (5.7)                          \\
3|4                        & 0.60 (13.5)                          & 1.02 (13.3)                          \\
4|5                        & 0.99 (22.3)                          & 1.58 (20.4)                          \\
\midrule
Residual deviance          & 691,240                             & 887,270                            \\
AIC                        & 691,256                             & 887,286                             \\ \bottomrule
\end{tabular}
\end{table}

Table \ref{tab:model_results_veh} shows the results for the model using vehicle deficiency as the treatment variable. This model differs from the previous two models in its focus on vehicle ownership rather than dwelling features. Vehicle deficiency, measured as fewer vehicles per household than adults, is found to increase the likelihood of delivery in the 2017-2019 period but decrease it in the 2019-2023 period. The interaction terms suggest that vehicle deficient households decreased their home delivery frequency relative in the 2017-2019 period but increased delivery frequency in the 2019-2023 period compared with other households.

\begin{table}[h]
\centering
\caption{Ordered probit model results for shopping frequency considering vehicle deficiency}
\label{tab:model_results_veh}
\begin{tabular}{@{}lcc@{}}
\toprule
\textbf{Variable}               & \textbf{2017-2019 Estimate (t-value)} & \textbf{2019-2023 Estimate (t-value)} \\ \midrule
Vehicle deficient               & 0.25 (54.0)                           & -0.19 (80.4)                          \\
(Year = $T_1$)                            & 0.15 (39.3)                           & 0.29 (161.5)                          \\
Vehicle deficient $\times$ (Year = $T_1$) & -0.38 (62.3)                          & 0.24 (79.1)                           \\
0|1                             & -0.53 (156.4)                         & -0.59 (414.5)                         \\
1|2                             & 0.23 (68.2)                           & 0.074 (53.1)                          \\
2|3                             & 0.68 (200.0)                          & 0.46 (324.6)                          \\
3|4                             & 0.97 (282.5)                          & 0.77 (525.5)                          \\
4|5                             & 1.30 (363.7)                          & 1.10 (719.1)                          \\ \midrule
Residual Deviance               & 2,103,605                             & 7,543,511                             \\
AIC                             & 2,103,621                             & 7,543,527                             \\ \bottomrule
\end{tabular}
\end{table}

\FloatBarrier

\section{Concluding Remarks}
There is ample evidence to suggest that e-shopping increased following the COVID-19 pandemic \cite{Shaw_Eschenbrenner_Baier_2022, Young_Soza‐Parra_Circella_2022, Bureau}. Factors that influence e-shopping demand include mid- to long-term household decisions (life events) such as household ownership, dwelling type, and vehicle ownership among other socio-demographic variables such as age, gender, and income. However, the impact of changes in life events and variation in the composition of e-shopping across demographic and dwelling strata is less clear. Such analyses require panel survey data which tracks individuals and household over time. However, such longitudinal datasets are rare and expensive to collect. In this study, we used a pseudo-panel DiD approach to address this research gap. This approach allows the analyst to approximate longitudinal effects using repeated cross-sectional survey data.

While we found an overall trend of increasing e-shopping over the period 2017-2023, shifts within the population were also observed via ordered probit models. Owners tended to  engage in more e-shopping in 2019 relative to 2017 but then decreased their e-shopping rate in relative terms by 2023. In contrast, apartment residents tended to receive fewer e-shopping deliveries than SFDTH residents across the analysis period, which tends to contradict the assumption of high e-shopping prevalence among apartment residents. As compared to other households, we find a decrease in e-shopping demand for vehicle deficient households in the 2017-2019 (pre-COVID) period; however, this behavior is reversed post-COVID. These results suggest that changes in life-events could influence e-shopping demand and this aspect is important to consider in e-shopping demand modeling. Neglecting this aspect could have policy implications. For instance, response to policies related to e-commerce may be gradual rather than immediate. 

This study is a first step towards approximating longitudinal effects using repeated cross-sectional survey data. Future studies could apply this methodology on survey data from other regions or countries and explore other life events. This study uses a limited set of variables, which could be expanded to include additional control variables. It may also be instructive to compare effects across delivery categories. For example, prepared food and grocery delivery patterns may differ across demographics strata. Pseudo panel construction will always remain a second-best alternative to \emph{true} panel data. However, it is a cost effective option particularly when one is limited to existing datasets. 

\section{Acknowledgments}
The authors appreciate PSRC making their travel diary openly available to the transportation community.

\section{Authors Contribution}
The authors confirm equal contribution to all aspects of the study conception and design, data collection, analysis, and paper writing.

\section{Conflict of Interest}
The authors declare no conflicts of interest with any other entities or researchers.

\bibliographystyle{trb}
\bibliography{references}
\end{document}